\documentclass[]{aa}
\titlerunning{Evidence for Binary Synchronization}

\usepackage[varg]{txfonts}  
\usepackage{graphicx}       
\usepackage{amsmath}        
\usepackage{xcolor}        
\usepackage{hyperref}     
\usepackage{placeins}
\usepackage{orcidlink}
\usepackage{float}

\hypersetup{
    colorlinks=true,         
    linkcolor=blue,       
    citecolor=blue,       
    urlcolor=blue      
}

\newcommand\gaia{\textit{Gaia }}

\newcommand\RGaia{$11\,500$}

\newcommand\Nvbroad{$3\,524\,677$}
\newcommand\NnummyEB{$2\,184\,477$}

\begin{document}

\title
{ Evidence for synchronisation of {\it Gaia} eclipsing binaries from their stellar rotational velocity}

\author{
    E. Hadad$^1$\thanks{Corresponding author; eyalhadad@mail.tau.ac.il},
     T. Mazeh$^1$\orcidlink{0000-0002-3569-3391},
    and S.  Faigler$^1$
}

\institute{
    School of Physics and Astronomy, Tel Aviv University, Tel Aviv, 6997801, Israel
}

\abstract{
The \gaia DR3 catalogue includes line-broadening measurements (\texttt{vbroad}) for $10\,387$ eclipsing binaries. In this study, we focus on a subset of $977$ short-period main-sequence sources with primary radii in the range of $ [1.25, 3]$ R$_\odot$, effective temperatures between $ 5\,600$ and $8\,000$ K, orbital periods from $0.3$ to $3$ days, \texttt{vbroad} measurements between $30$ and $300$ km/s, eclipse depth ratios below $0.7$ and $|e\cos\omega| < 0.1$. Our analysis reveals a clear inverse correlation between the derived \texttt{vbroad} values and the orbital periods of these systems, consistent with tidal synchronisation and spin–orbit alignment.

We further compare the \gaia \texttt{vbroad} results with the expected rotational velocities of the primaries of these binaries, given the radius estimate of these stars. 

We find that the \texttt{vbroad} measurements are generally consistent with the expected rotational broadening, except for a systematic lower factor of approximately
10\%.
}


\maketitle

\section{Introduction}
The \gaia space mission 
\cite{Gaia16}
is equipped with a spectrometer featuring an intermediate resolving power of R $\approx$ \RGaia, covering a wavelength range of $846$ to $870$ nm, whose primary objective is to measure the radial velocities (RVs) of bright sources 
\citep{cropper18}.\footnote{\href{https://gea.esac.esa.int/archive/documentation/GDR3/index.html}{https://gea.esac.esa.int/archive/documentation/GDR3/index.html}}
Additionally, the spectroscopic pipeline \citep{Sartoretti18}
provides a line-broadening
parameter that characterises the broadening of the absorption lines relative to the pertinent template \citep[]{Fremat23}, yielding \texttt{vbroad} measurements for \Nvbroad~sources.\footnote{\href{https://gea.esac.esa.int/archive/documentation/GDR3/Gaia_archive/chap_datamodel/sec_dm_main_source_catalogue/ssec_dm_gaia_source.html\#gaia_source-:~:text=vbroad\%20\%3A\%20Spectral\%20line,\%5D\%20)}{https://gea.esac.esa.int/archive/documentation/GDR3/ \\ Gaia\_archive/chap\_datamodel/sec\_dm\_main\_source\_catalogue/ \\ ssec\_dm\_gaia\_source.html\#gaia\_source-:~:text=vbroad\%20\%3A\%20Spectral\%20line,\%5D\%20)}}  \citet[]{Fremat23} compared their results with other large spectroscopic surveys --- RAVE DR6
\citep{RAVE20}, GALAH DR3 \citep{Buder2021}, APOGEE DR16 \citep{Jonsson20}, and
LAMOST DR6 \citep[]{Xiang22}, and showed reasonable agreement for \texttt{vbroad} larger than $10$ km/s.

Despite the relatively low resolution of the spectrograph, the measured \texttt{vbroad} can be used to derive the stellar rotational broadening, as discussed by \citet[]{Fremat23}. 
In a previous paper \citep[][]{hadad25} we showed that some unresolved wide binaries with bright enough secondaries, identified by their \gaia brightness excess 
relative to their 
BP-RP colour, display relatively large \texttt{vbroad} values. We interpreted this effect as the contribution of the lines of the secondaries of these systems, which are shifted by their relative orbital motion. 

For this paper we considered the derived \texttt{vbroad} for stars included in the large sample of \gaia main-sequence (MS) short-period eclipsing binaries (EBs) identified by \cite{mowlavi23}.\footnote{\href{https://gea.esac.esa.int/archive/documentation/GDR3/Gaia_archive/chap_datamodel/sec_dm_variability_tables/ssec_dm_vari_eclipsing_binary.html}{https://gea.esac.esa.int/archive/documentation/GDR3\\/Gaia\_archive/chap\_datamodel/sec\_dm\_variability\_tables/\\ssec\_dm\_vari\_eclipsing\_binary.html}} 
The two \gaia catalogues bring a unique opportunity to compare the orbital periods of a large sample of EBs, as derived from the \gaia light curves, with the stellar rotation periods of their primaries, as reflected by their \texttt{vbroad}.

Many recent studies have derived rotational periods from observed stellar photometric modulation \citep[e.g.][]{McQuillan13b, McQuillan13a, mcquillan14, aigrain15,breton21}, under the assumption that they reflect some imperfect symmetry of the stellar surface. 
These methods have been instrumental in the study of synchronisation in binary star systems. 
\cite{lurie17} analysed $2\,278$ EBs spanning spectral types A through M, and reported that 79\% of EBs with orbital periods shorter than $10$ days are synchronised. 
 
However, the photometric modulations are not necessarily coherent, as their amplitudes and phases can change over time. Furthermore, these variations can be masked by other stellar photometric modulations. Therefore, it is important to explore alternative methods for deriving stellar rotation periods, leveraging the available large spectroscopic samples. 

Few studies have published large samples of synchronised binaries based on measuring the rotational broadening of their spectra ($V\sin i$). \cite{Levato1976} examined $V\sin i$ measurements for $122$ known binaries and concluded that synchronisation is evident in all systems with orbital periods shorter than $3$ days, and can be found in periods of up to $6$ days. More recently, \cite{Lennon2024} analysed the rotational velocities and periods of $73$ B-type SBs in the LMC, which provided strong evidence for synchronisation in all systems with periods shorter than $3$ days. 

For our work, we considered the synchronisation of $977$ short-period EBs, an achievement made possible by the size and richness of the \gaia \texttt{vbroad} and EBs catalogues.
We cross-matched the two catalogues and showed that the \texttt{vbroad} measurements of the short-period binaries are inversely proportional to their orbital periods, as expected for synchronised and aligned systems. 

To show that this is indeed expected, we note that the stellar equatorial rotational velocity, $V_*$, is

\begin{equation}
\label{eq: synchronous rotation equation}
   V_*=\frac{2 \pi R_{*}} {P_*}\, ,
\end{equation}
where $R_*$ and $P_*$ represent the stellar radius and rotational period of the primary star, respectively. 

We assume that 
\begin{itemize}
    \item $\texttt{vbroad} \simeq V_*\sin{i_*}$, where ${i_*}$ denotes the inclination of the stellar rotation axis (\texttt{vbroad} is  not affected by other broadening effects),
    \item $\sin{i_*} \simeq 
            \sin{i_{\rm orb}}$ (alignment of the stellar rotation with the orbital motion),
     \item $\sin{i_{\rm orb}} \simeq 1$ (the binary is eclipsing),  
     \item $P_*\simeq P_{\rm orb}$ (synchronisation of the stellar and orbital motions).
    
\end{itemize}
Then, 
\begin{equation}
\label{eq:vbroad_period}
\texttt{vbroad}\simeq\frac{2 \pi R_{*}}{\texttt{period}} \, ,
\end{equation}
where \texttt{period} is $P_{\rm orb}$ of the EB.

We expect the effect of synchronisation to be more apparent for binaries with faint companions. This is so because, as seen in our previous paper, the  $\texttt{vbroad}$ measurements for binaries with relatively bright secondaries are contaminated by the contribution of the secondary. The flux ratio of the two components of each EB is reflected by the light curve secondary to primary eclipse depth ratio, derived by the \citet{mowlavi23} analysis. We therefore divided the cross-matched sample into binaries with low and high depth ratios. 

Stellar synchronisation is generally expected to occur concurrently with orbital circularisation. Equation \ref{eq:vbroad_period} applies strictly to circular orbits. To account for this, we utilised the eclipse phase information provided in the EB catalogue to identify and exclude binaries with eclipse separations that deviate significantly from half an orbital period, which is indicative of orbital eccentricity.

We demonstrated that the \texttt{vbroad} values of the selected sample
are inversely proportional to the orbital periods
derived from their \gaia lightcurves. 

Section~\ref{sec: selecting the sample} describes the filtering process applied to the sample, Sect.~\ref{sec: results} quantifies the observed trend, and Sect.~\ref{sec: discussion} discusses our results.

\section{Selecting the sample}
\label{sec: selecting the sample}

\gaia Data Release $3$ (DR$3$) includes a catalogue of \NnummyEB~eclipsing binary candidates \citep[]{mowlavi23}, of which
$10\,387$ sources have a \texttt{vbroad} measurement. 

We chose to restrict our analysis to MS stars. This was done by constructing a colour-magnitude diagram (CMD) for stars with small parallax uncertainty. Specifically, we only included sources with ${\varpi} / {\sigma_{\varpi}} > 10$, 
where $\varpi$ and $\sigma_{\varpi}$ are the \gaia parallax and its uncertainty,
yielding a sample of $9\,911$ stars. Among these, $7\,856$ sources are provided with extinction parameters in \gaia DR$3$, and they were used in plotting the CMD location.

\begin{figure}
\includegraphics[width=\columnwidth, height=8cm]{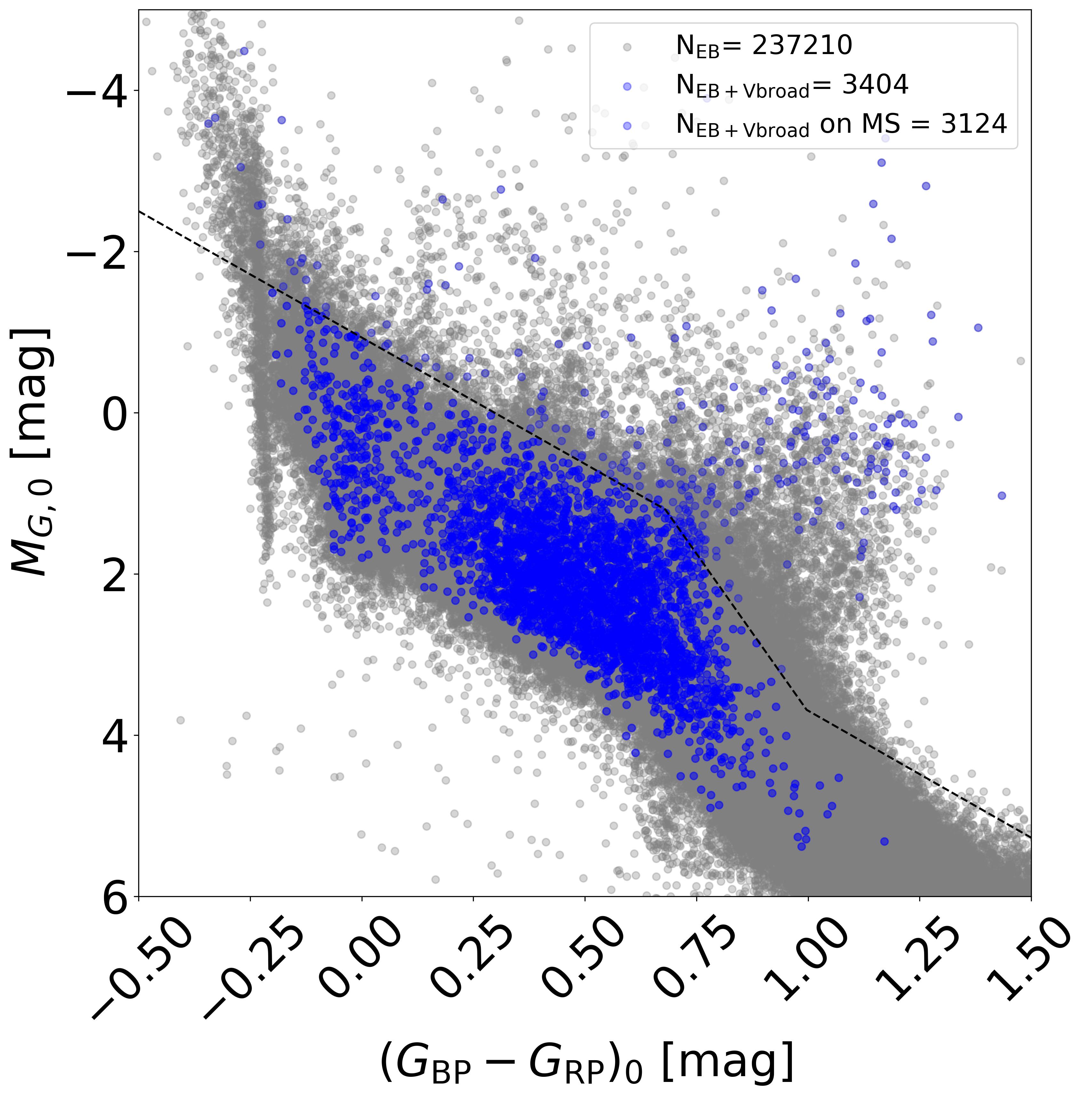}
\caption{Selecting a MS subsample of the EB catalogue. In grey: all \cite{mowlavi23} catalogue sources with ${\varpi} 
 / {\sigma_{\varpi}} > 10$, that have extinction parameters. In blue: $3\,404$ sources with \texttt{vbroad}, for which \texttt{vbroad} / $\sigma_{\mathrm{vbroad}} > 2$; the black line delineates the upper boundary of the main sequence---$3\,124$ sources are below the line. The black line is defined by the vertices: $(-0.5, -2.5), (0.68, 1.2), (0.99, 3.68), (1.5, 5.28)$.}
    \label{fig:nummy on MS}
\end{figure}

Furthermore, to ensure reliable broadening velocity measurements, we imposed a significance criterion of $\texttt{vbroad} / \sigma_{\texttt{vbroad}} > 2$, which reduced the sample to $3\,404$ sources. These stars are presented in Fig.~\ref{fig:nummy on MS}, with $3\,124$ identified as MS stars based on their position in the CMD.

\begin{table}[htb]
    \centering
    \small
    \begin{tabular}{p{5cm}r}
            \hline
        Criterion & Number of Sources\\
                \hline
        EBs with \texttt{vbroad} & $10\,387$\\
        $\varpi / \sigma_{\varpi} > 10$ & $9\,911$\\
        With extinction coefficients & $7\,856$\\
        $\texttt{vbroad} / \sigma_{\texttt{vbroad}} > 2$ & $3\,404$\\
        On the MS & $3\,124$\\
        $1.25 \leq $ Radius $\leq 3$ R$_\odot$; $5\,600 \leq$ Temperature $\leq 8\,000$ K &  $2\,260$\\
        $30 \leq \texttt{vbroad} \leq 300$ km/s; $0.3 \leq \texttt{period} \leq 3$ days & $2\,082$\\
        Depth ratio $< 0.7$ & $1\,050$\\
        $|e\cos\omega|$ $< 0.1$ & $977$\\
                \hline
    \end{tabular}
    \caption{Summary of the filtering process.}
    \label{tab:filtering process}
\end{table}

To reduce the radii variance of the sample  
(see Eq.~(\ref{eq:vbroad_period})), we constrained it to primaries with radii in the range $[1.25, 3]$ R$_{\odot }$
and effective temperatures in the range $[5800, 8000]$ K, 
using \texttt{radius\_gspphot} and \texttt{teff\_gspphot} from \textit{Gaia}.\footnote{\href{https://gea.esac.esa.int/archive/documentation/GDR3/Gaia_archive/chap_datamodel/sec_dm_astrophysical_parameter_tables/ssec_dm_astrophysical_parameters.html\#astrophysical_parameters-}{\url{https://gea.esac.esa.int/archive/documentation/GDR3/Gaia_archive/chap_datamodel/sec_dm_astrophysical_parameter_tables/ssec_dm_astrophysical_parameters.html\#astrophysical_parameters-}}}
Applying this selection criterion resulted in $2\,260$ sources.

\begin{figure}[!h]
	\includegraphics[width=\columnwidth, height=8cm]{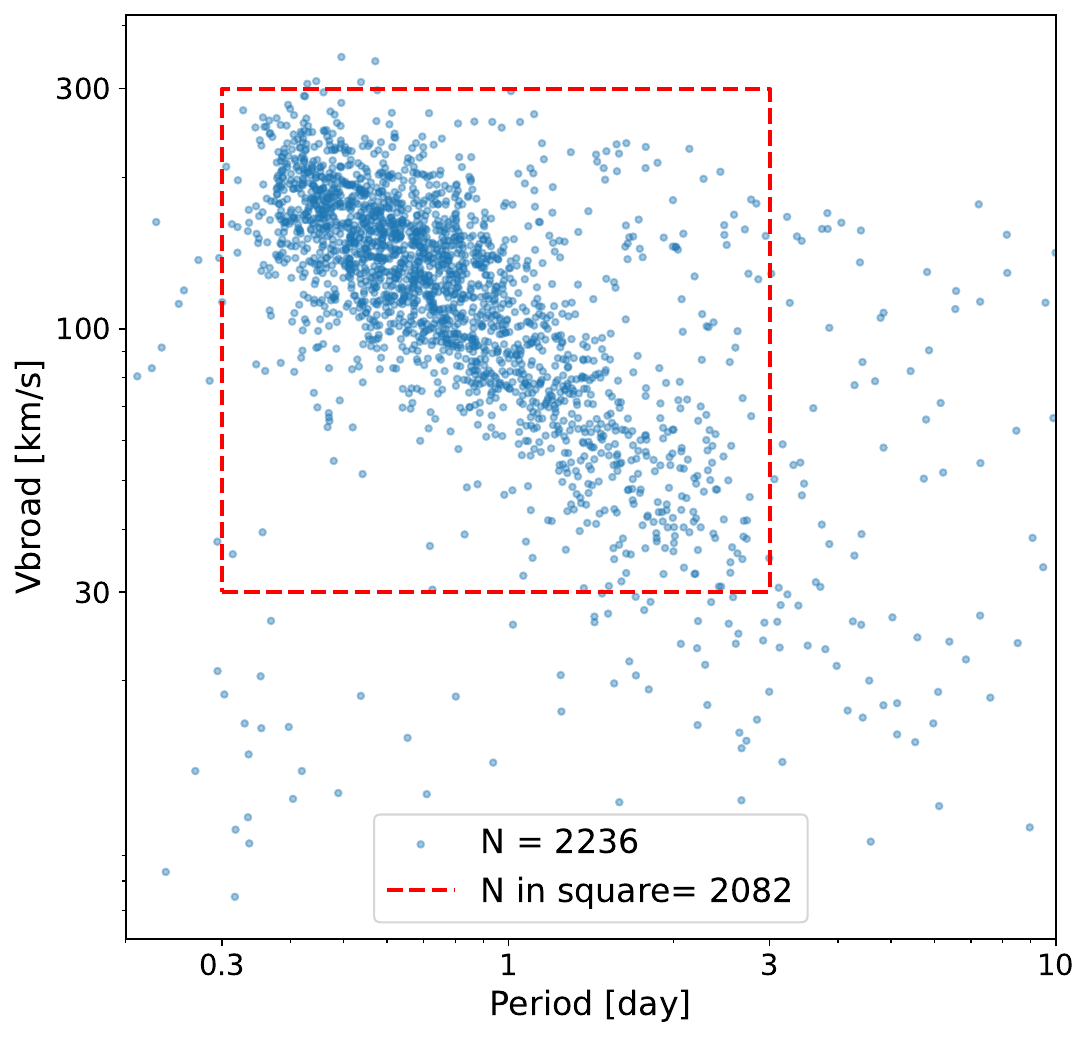}
\caption{\texttt{vbroad} vs \texttt{period} for the MS eclipsing binary
population of Fig~\ref{fig:nummy on MS}, limited by $\texttt{vbroad}>10$ km/s and $\texttt{period} \in [0.2,10]$ days. The red square marks our analysis region ($2\,082$ sources).}
    \label{fig:nummy vbroad v period on MS}
\end{figure}

Further focusing on sources with $\texttt{vbroad}>10$ km/s and orbital \texttt{periods} in the range $[0.2,10]$ days left us with $2\,236$ sources. These are plotted in Fig.~\ref{fig:nummy vbroad v period on MS}, which shows a clear linear correlation in log scale, where \texttt{vbroad} is inversely proportional to the orbital period, as expected for a synchronised sample.

\begin{figure}[!h]
    \centering
\includegraphics[width=\columnwidth]{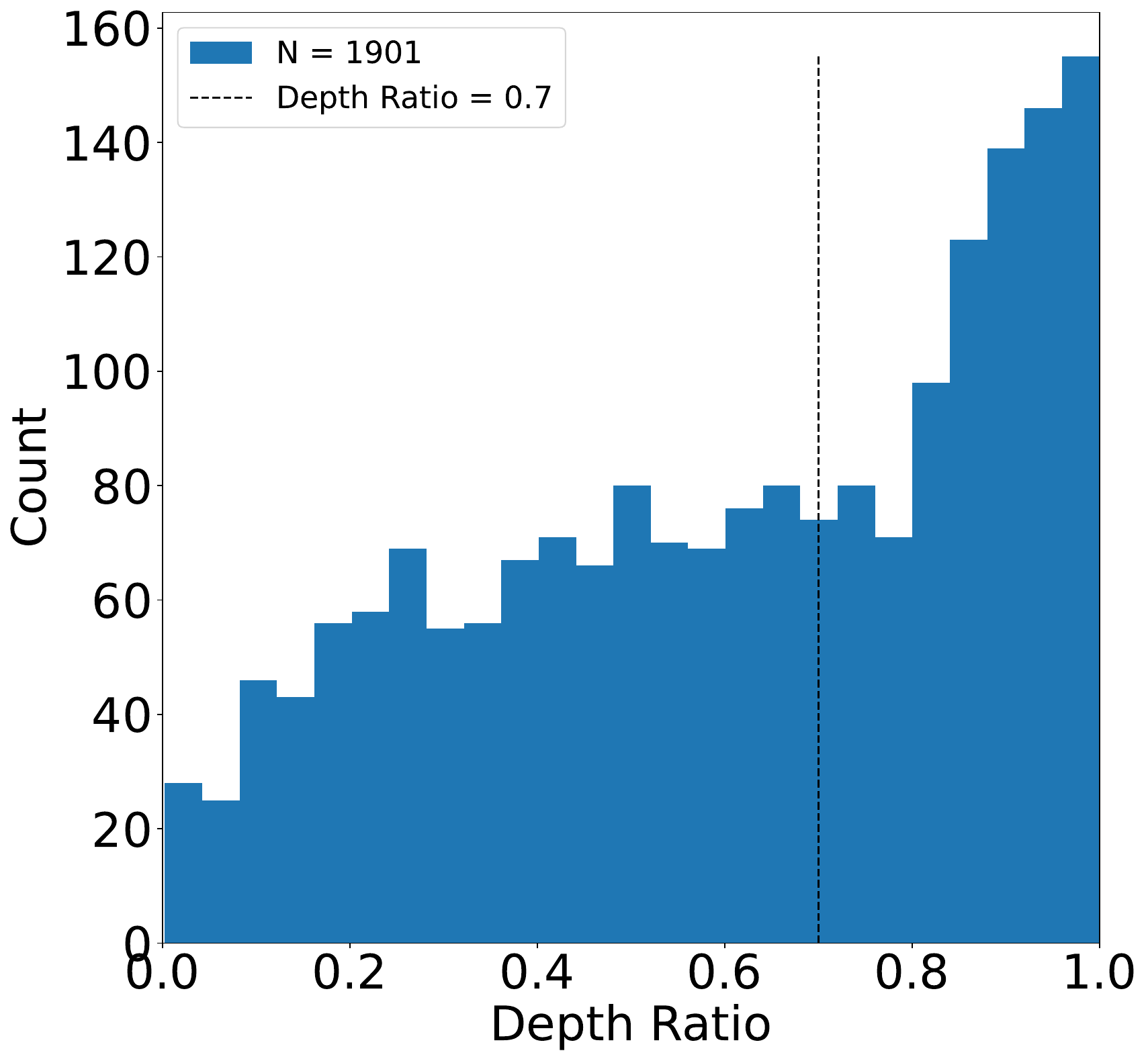}
    \caption{Depth ratio histogram showing the $0.7$ threshold applied to select $1\,050$ systems with faint secondaries for further analysis.}
    \label{fig:depth ratio histogram}
\end{figure}
Most of the binaries ($2\,082$ sources) are within a \texttt{vbroad} range of $30$--$300$ km/s and a \texttt{period} range of $0.3$--$3$ days (red square). Somewhat arbitrarily, we chose to analyse the correlation of this restricted sample.

In addition, we focused on EBs with low depth ratios. Around 94$\%$ of the sources from the original EB sample were fitted by \cite{mowlavi23} with a two-Gaussian model, obtaining two depth parameters: \texttt{derived\_primary\_ecl\_depth} and \texttt{derived\_secondary\_ecl\_depth}. We calculated the EB secondary-to-primary depth ratio using these two parameters, which reduced the sample to $1\,901$ sources.

Based on the depth ratio histogram (Fig.~\ref{fig:depth ratio histogram}), we restricted our analysis to EBs with depth ratios below $0.7$, selecting systems with faint secondaries to ensure more reliable \texttt{vbroad} measurements. This yielded $1\,050$ sources. 

Finally, we restricted our sample to systems with low eccentricities. To this end, we made use of the catalogue’s eclipse phase estimates (\texttt{derived\_primary\_ecl\_phase} and \texttt{derived\_secondary\_ecl\_phase}). We used the small $e$ approximation
$e\cos\omega \approx \frac{\pi}{2}(\phi_2-\phi_1-0.5)$,
where $\phi_{1,2}$ are the two eclipses phases. Applying a threshold of $|e\cos\omega| < 0.1$ further reduced the sample to $977$ systems. A summary of the filtering process is provided in Table \ref{tab:filtering process}.

\section{Analysis}

\label{sec: results}
We developed a tailored Markov chain Monte Carlo
(MCMC) algorithm, similar to the one developed by \cite{bashi23}, that searches for a probability density function $\mathcal{F}_{\rm PDF}$ that best describes the distribution of the points 
within the 2D region bounded by the red lines in Fig.~\ref{fig:nummy vbroad v period on MS}.
The algorithm fits the data with a $2$D Gaussian function, which is rotated relative to the diagram axes. A full description of the algorithm is provided in Appendix \ref{app:mcmc algorithm}.

\begin{figure}[]
\centering
\includegraphics[width=\columnwidth]{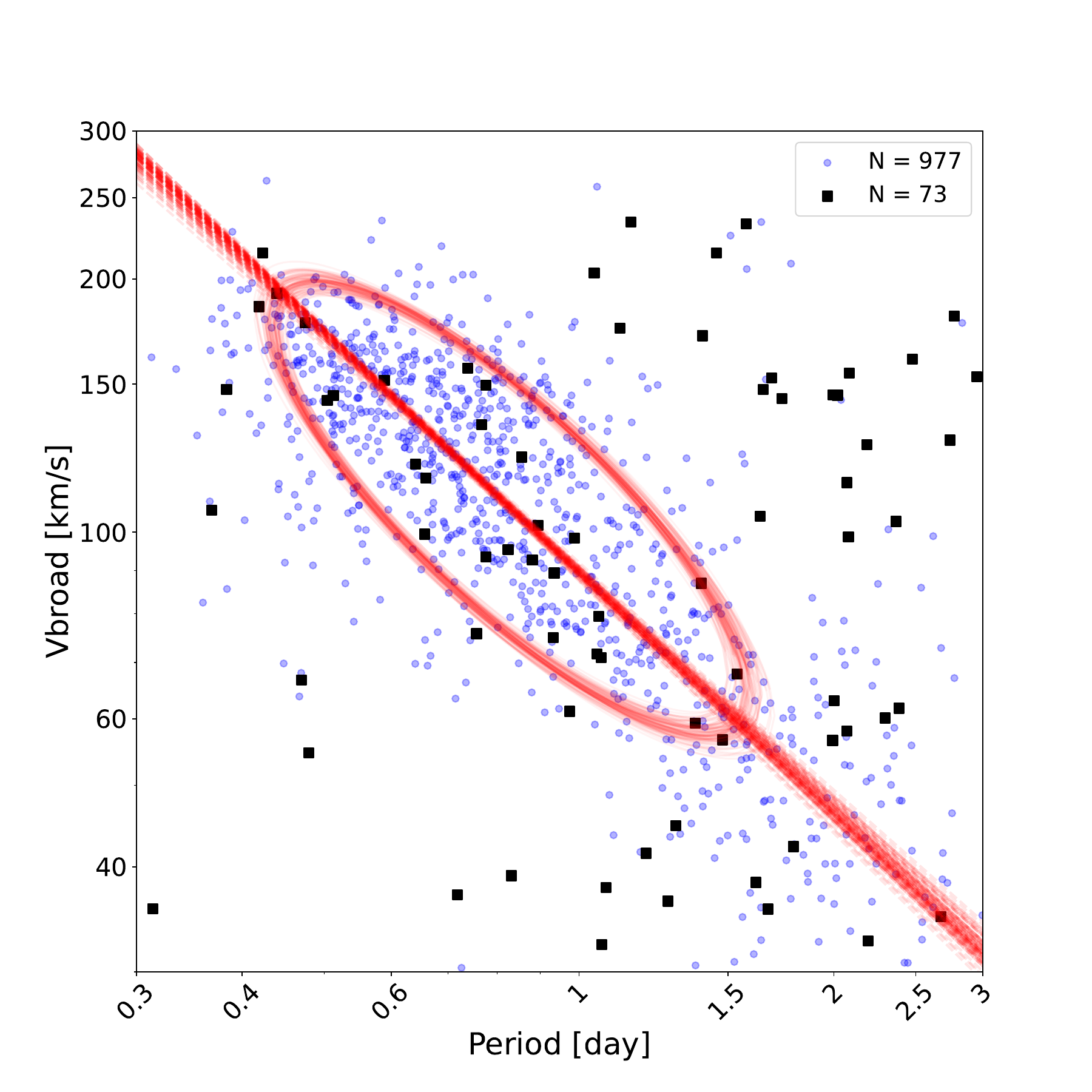}
\caption{
Rotated 2D Gaussian representing the distribution of EBs in the 
($\texttt{vbroad}$, \texttt{period}) plane. Line equation: $\log V=A \log P+B$, where $A = -0.948 \pm 0.026$ and $B = 1.952 \pm 0.005$. The black squares mark systems with $|e\cos\omega| > 0.1$, which are not fitted by the algorithm. The plot shows $100$ parameter sets sampled from the MCMC chains after convergence. The red lines represent the Gaussian's major-axis orientation, 
and the red ellipses show the distribution $3\sigma$ region. A full derivation is provided in Appendix \ref{app:red line fit}.}
    \label{fig: Gaussian fit}
\end{figure}

The results are presented in Fig.~\ref{fig: Gaussian fit}, which plots \texttt{vbroad} against \texttt{period} on a logarithmic scale. The best-fit parameters of the model are listed in Table \ref{tab:1-s values}.
The slope of the 
$2$D Gaussian function is $-0.948 \pm 0.026$. This linear trend, measured in log space, lies within $2\sigma$ of $-1$, consistent with synchronous rotation (Eq.~(\ref{eq:vbroad_period})). 

\begin{figure}[]
\centering
\includegraphics[width=\columnwidth]{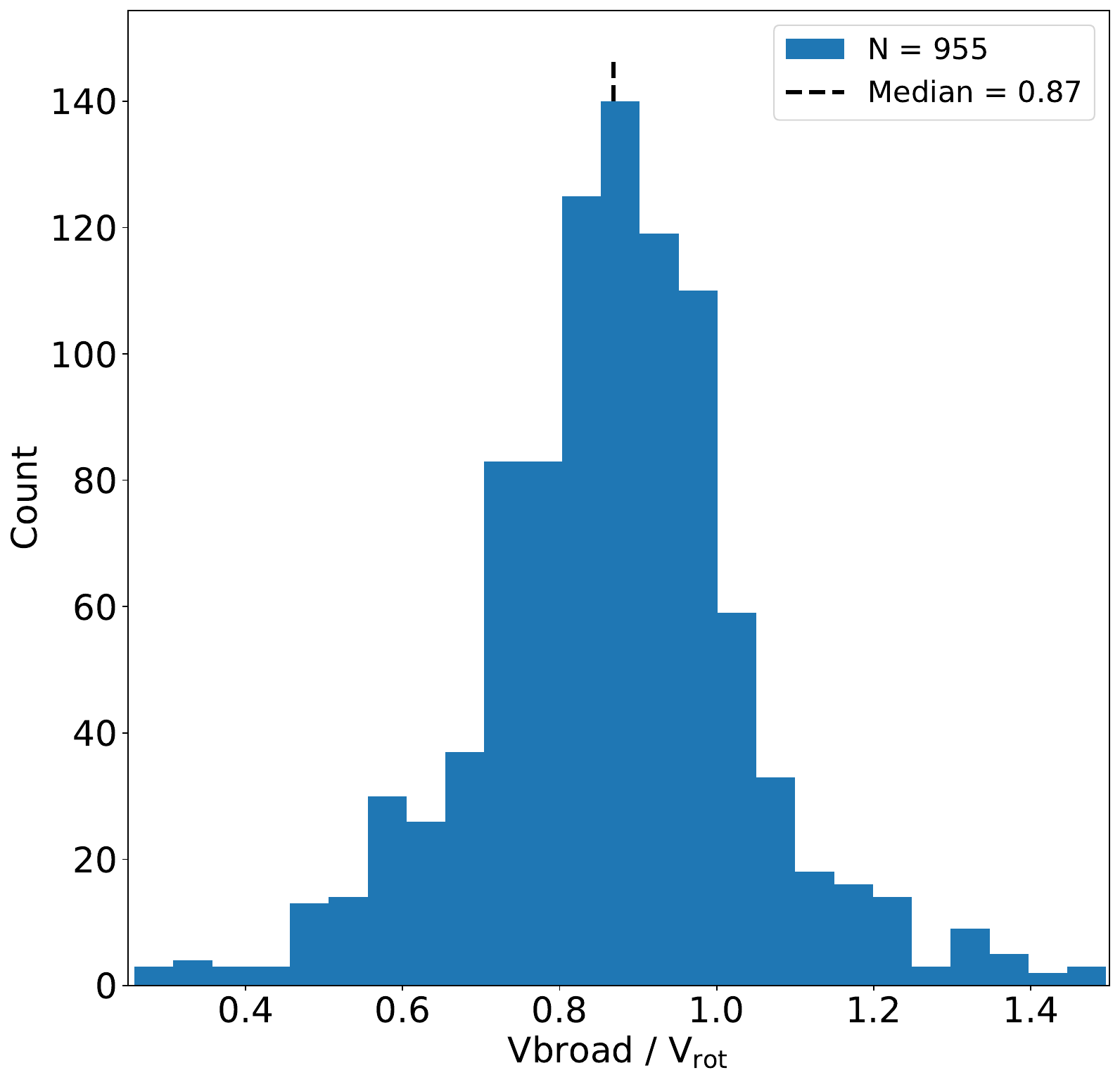}
\caption{\texttt{vbroad}/$\mathrm{v_{rot}}$ for the synchronised binaries. The sample is limited to sources with $\texttt{vbroad} / \mathrm{v_{rot}}$ in the range of $[0.25, 1.5]$ for visual clarity. The black vertical line marks the median of the sample. }
\label{fig: vbroad vs vrot}
\end{figure}

Following Eq.~(\ref{eq: synchronous rotation equation}), while assuming that the short-period EBs are synchronised and aligned, we used the {\it Gaia}-derived radii and the EB \texttt{period} to estimate the expected rotational velocity of the primary star, denoted $\mathrm{v_{rot}}$. We then constructed a histogram of the ratio $\texttt{vbroad} / \mathrm{v_{rot}}$, shown in Fig.~\ref{fig: vbroad vs vrot}. The median value of the distribution is $0.87$, indicating that \texttt{vbroad} is broadly consistent with $\mathrm{v_{rot}}$, though it exhibits a systematic lower offset of approximately $\sim10\%$.

\section{Discussion}
\label{sec: discussion}

The analysis of the \textit{Gaia} \texttt{vbroad} derived for a sample of $977$ short-period EBs yielded two results.

\begin{itemize}
    \item The \texttt{vbroad} values are inversely proportional to the orbital period of the EBs.
    
    \item{Assuming the binaries of the sample are synchronised and aligned, the expected equatorial velocities of their primaries are, up to about $10\%$, similar to their derived \texttt{vbroad}}.
\end{itemize}

These results suggest two things:

\begin{itemize}
    \item The short-period binaries are synchronised and aligned. 
   
    \item The derived \texttt{vbroad} represents the actual stellar equatorial velocity, with a systematic offset of $\sim10\%$ lower.
\end{itemize}

Several studies have investigated the accuracy and robustness of deriving the stellar rotation velocity from spectral line broadening.
For example, \cite{royer02a,royer02b} conducted a large study of B8–F2-type stars across the southern and the northern skies. They reported that their $V\sin i$ measurements were systematically $\sim 10\%$ higher than those reported by
\cite{slettebak75} (see also \cite{slttebak55}), probably due to the 
presence of binaries in the sample used by the early study. In contrast, our analysis compared the spectroscopically derived stellar rotation with those derived from the orbital periods obtained by photometry of EBs
\citep[see below][]{Levato1976,Lennon2024}.

The approximately $10\%$ lower \texttt{vbroad} we found may result from several factors. It could be the result of the inclination of the stellar rotational axis.
However, we expect this factor to be small, as the systems are eclipsing, so their orbital inclinations are close to $90^\circ$, and we expect our systems to be aligned. 

Another contributing factor could be an overestimation of the primary stellar radius due to the luminosity of the secondary component. Nevertheless, this is unlikely to fully explain the discrepancy,
as we 
deliberately selected binaries with relatively faint secondaries, most of which contribute much less than $\sim 20\%$ of the total system luminosity. Such a small contribution would be insufficient to account for the $\sim 10\%$ increase in stellar radius (and, consequently, equatorial velocity) required to explain the observed discrepancy. 

Therefore, the systematic offset may be attributed to residual biases in the analysis or underestimation of other line-broadening mechanisms.

Apart from the $10\%$ systematic shift, the \gaia \texttt{vbroad} catalogue, and the catalogues of $V\sin i$ from other large spectroscopic surveys, such as GALAH \citep{Buder2021} and APOGEE \citep{Jonsson20}, can be used to estimate the stellar rotation period of large samples, 
albeit with lower accuracy due to their reliance on knowledge of stellar radius and inclination.

Observational evidence for synchronisation and alignment of samples of short-period binaries is crucial to our understanding of the long-term tidal interaction acting in close binaries  
\citep[e.g.,][]{hut1981, Zahn1989, ZahnBouchet89}. In particular, the period limits between synchronised and non-synchronised binaries \citep[e.g.,][]{mathieu88} and their dependence on spectral type 
\citep[e.g.,][]{Giuricin1984,khaliullin10}
and mass ratio are essential. Unfortunately, the present \gaia EB catalogue is limited to short-period binaries, and the evidence for synchronisation of binaries with periods shorter than three days could be expected.  However, the next \gaia data release, DR4, 
which will incorporate $66$ months of observations compared to the $34$ months included in DR3,
will have longer-period EBs, enabling further study of tidal interaction.

\begin{acknowledgements}

We are deeply grateful to the referee, Dr. Nami Mowlavi, for his thorough review of the manuscript and for his insightful comments and suggestions, which greatly enhanced the clarity, structure, and scientific rigour of this work. His expertise and attention to detail were particularly valuable in refining the selection process to achieve more robust results. We deeply appreciate the time and effort he dedicated to the review process.
We are also grateful to the \gaia CU6 and CU7 teams for releasing the \texttt{vbroad} and EB catalogues, which enabled this study through their extensive, high-quality data. 
TM is grateful for the support of Israel
Ministry of Science Grant no~3-18140.
This work has made use of data from the European Space Agency (ESA) mission Gaia (https://www.cosmos.esa.int/gaia), processed by the Gaia Data Processing and Analysis Consortium (DPAC, https://www.cosmos.esa.int/web/gaia/dpac/consortium). Funding for the DPAC has been provided by national institutions, in particular those participating in the Gaia Multilateral Agreement.
\end{acknowledgements}

\bibliographystyle{aa}
\bibliography{main} 

\clearpage
\begin{appendix}
\section{}
\label{app:mcmc algorithm}

For each parameter set of the distribution $\mathcal{F}_{\rm PDF}$---$\eta$, 
we derive the likelihood of $N$-binaries sample, 
as
\begin{equation}
\mathcal{L} = \prod_{i=1}^{N} \mathcal{F}_{\mathrm{PDF}}
(\texttt{vbroad}_i, \texttt{period}_i; \eta) \ .
\label{eq:likelihood}
\end{equation}

where $\texttt{vbroad}_i$ and $\texttt{period}_i$ are the \texttt{vbroad} and \texttt{period} of the $i$-th binary, given in km/s and days, respectively. We then find the parameter set $\eta$
that maximises the likelihood of the sample.

The model function consists of a two-dimensional Gaussian combined with a uniform background density. 
It is characterised by the Gaussian centre coordinates $(\log\texttt{p}_0, \log\texttt{v}_0)$: its angle, clockwise from the x-axis, $\theta$; the standard deviations (in log space): $\sigma_x, \sigma_y$, and the peak density at the centre, $D$. 
The Gaussian is added to a rectangle with constant background density $D$. 
Therefore, given a set of parameters $\eta$, each point $( \texttt{vbroad}_i, \texttt{period}_i)$ is assigned a probability of
\begin{equation}
\label{eq:probability value}
   \mathcal{F}_{\mathrm{PDF}}
(vbroad_i, period_i; \eta) = D e^{-(aP^2 + bPV + cV^2)} + C \ ,
\end{equation}
where
\begin{equation*}
P = \log\texttt{period}_i - \log\texttt{p}_0 \ ,
\end{equation*}
\begin{equation*}
V = \log\texttt{vbroad}_i - \log\texttt{v}_0 \ ,
\end{equation*}
and the $a, b$ and $c$ coefficients are
\begin{equation*}
\label{eq:a coefficients}
a = \frac{\cos^2\theta}{2\sigma_x^2} + \frac{\sin^2\theta}{2\sigma_y^2} \ ,    
\end{equation*}
\begin{equation}
\label{eq:b coefficients}
    b = -\sin2\theta (\frac{1}{4\sigma_x^2} - \frac{1}{4\sigma_y^2}) \ ,   
\end{equation}
\begin{equation*}
\label{eq:c coefficients}
    c = \frac{\sin^2\theta}{2\sigma_x^2} + \frac{\cos^2\theta}{2\sigma_y^2} \ .   
\end{equation*}
The domain area in log space is 1, so $C$ is derived from the $\mathcal{F}_{\rm PDF}$ normalisation
\begin{equation}
\label{eq:normzalization}
    D 2\pi\sigma_x\sigma_y + C = 1 \ .
\end{equation}
 
\section{}
\label{app:red line fit}

For a given parameter set $\eta$, a red line in Fig.~\ref{fig: Gaussian fit} is calculated as
\begin{equation}
    y=Ax+B \ , 
\end{equation}
for 
\begin{equation}
    A=-\tan\theta \ , 
\end{equation}
and 
\begin{equation}
    B=\log\texttt{v}_0 + \tan\theta \log\texttt{p}_0 \ . 
\end{equation}
The red ellipse's left and right focal points are
\begin{equation}
    (x_L, y_L)=(\log\texttt{p}_0-d \cos\theta, \log\texttt{v}_0+d \sin\theta) \ , 
\end{equation}
\begin{equation}
    (x_R, y_R)=(\log\texttt{p}_0+d \cos\theta, \log\texttt{v}_0-d \sin\theta) \ , 
\end{equation}
for
\begin{equation}
    d=\sqrt{|(3\sigma_x)^2-(3\sigma_y)^2|} \ .
\end{equation}
Then, the ellipse is the set of points $(x', y')$ which satisfies: 
\begin{equation}
    \sqrt{(x'-x_L) + (y'-y_L)} + \sqrt{(x'-x_R) + (y'-y_R)} = 2 max(3\sigma_x,3\sigma_y) \ . 
\end{equation}

\FloatBarrier

\begin{table}
\caption{
Figure \ref{fig: Gaussian fit} fit parameters:\
\label{tab:1-s values}}
\centering
\begin{tabular}{lc}
\hline
& Fitted Values\\
\hline
$\texttt{p}_0$ [day] & $0.828\pm0.013$\\
$\texttt{v}_0$ [km/s] & $107.0568\pm1.655$\\
$\theta$ [deg] & $43.458\pm0.779$\\
$\sigma_x$ & $0.248\pm0.006$\\
$\sigma_y$ & $0.076\pm0.003$\\
$D$ & $7.7525\pm0.361$\\
$A$ & $-0.948\pm0.026$\\
$B$ & $1.952\pm0.005$\\

\hline
\end{tabular}
\end{table}

\end{appendix}

\end{document}